%% file: main.tex
\documentclass[sigconf,10pt]{acmart}
\AtBeginDocument{%
  \providecommand\BibTeX{{%
    \normalfont B\kern-0.5em{\scshape i\kern-0.25em b}\kern-0.8em\TeX}}}
\usepackage{macros}
\usepackage{subcaption}
\usepackage{color}
\usepackage{amsmath}
\usepackage{subcaption}
\usepackage{graphicx}
\usepackage{tabularx}
\usepackage{array,multirow,rotating}
\usepackage{algorithm}
\usepackage{algpseudocode}
\usepackage{amsmath}
\usepackage{tikz}
\usepackage{adjustbox}
\usepackage{float}

\usepackage{mathtools}
\usepackage{lipsum}

\usepackage{mathtools}
\usepackage{tablefootnote} 
\usepackage{soul}
\usepackage{footnote}
\usepackage[toc,page]{appendix} 
\usepackage[]{hyperref}

\usepackage[inline]{enumitem}

\usepackage{wrapfig}
\usepackage{soul}

 \sloppypar
\def\footnoterule{\kern-3pt
  \hrule \kern 2.6pt} 

\setcopyright{none}
\renewcommand\footnotetextcopyrightpermission[1]{}
\begin{document}

\title{Synergistic and Efficient Edge-Host Communication for Energy Harvesting Wireless Sensor Networks}

\author{Cyan Subhra Mishra, Jack Sampson, Mahmut Taylan Kandmeir, Vijaykrishnan Narayanan, Chita R Das\\
\{cyan, jms1257, mtk2, vijaykrishnan.narayanan, cxd12\}@psu.edu\\
The Pennsylvania State University
}


\def\footnoterule{\kern-3pt
  \hrule \kern 2.6pt} 

\begin{abstract}
\input{src/00abstract}
\end{abstract}
\maketitle
\pagestyle{plain}

\section{Introduction}
\label{sec:introduction}
\input{src/01Introduction}

\section{Background and Motivation}
\label{sec:bg-rw}
\input{src/02BG-RW}

\section{Design Space Exploration}
\label{sec:DSE}
\input{src/03DesignSpaceExp}

\section{Design Implementation of \emph{Seeker}}
\label{sec:4}
\input{src/04Seeker}

\section{Evaluation and Results}
\label{sec:evaluation}
\input{src/05Evaluation}

\section{Conclusion}
\label{sec:conclusion}
\input{src/06Conclusion}
\bibliographystyle{ACM-Reference-Format}
\bibliography{references}

\clearpage
\appendix
\section{Appendix}
\label{sec:appendix}
\input{src/Appendix}
\section{Appendix}

\end{document}

%% file: src/00abstract.tex
There is an increasing demand for intelligent processing on ultra-low-power internet of things (IoT) device. Recent works have shown substantial efficiency boosts by executing inferences directly on the IoT device (node) rather than transmitting data. However, the computation and power demands of Deep Neural Network (DNN)-based inference pose significant challenges in an energy-harvesting wireless sensor network (EH-WSN). Moreover, these tasks often require responses from multiple physically distributed EH sensor nodes, which impose crucial system optimization challenges in addition to per-node constraints. To address these challenges, we propose \emph{Seeker}, a hardware-software co-design approach for increasing on-sensor computation, reducing communication volume, and maximizing inference completion, without violating the quality of service, in EH-WSNs coordinated by a mobile device. \emph{Seeker} uses a \emph{store-and-execute} approach to complete a subset of inferences on the EH sensor node, reducing communication with the mobile host. Further, for those inferences unfinished because of the  harvested energy constraints, it leverages task-aware coreset construction to efficiently communicate compact features to the host device. We evaluate  \emph{Seeker} for human activity recognition, as well as predictive maintenance and show $\approx 8.9\times$ reduction in communication data volume with $86.8\%$ accuracy, surpassing the $81.2\%$ accuracy of the state-of-the-art.



%% file: src/01Introduction.tex
Innovations in low-power computing, artificial intelligence, and communication technologies have given rise to the generation of intelligently connected devices that constitute the Internet of Things (IoT).  
Wireless sensor networks (WSNs), one of the prominent classes of IoT deployments, is currently dominating and expected to be pervasive impacting many application spaces~\cite{Cisco-report} including, but not limited to, body area network~\cite{Origin,batteryfree}, industrial monitoring~\cite{industry4}, predictive maintenance~\cite{industry40}, commercial satellites\cite{oribitalEH1} and smart farming~\cite{smartfarmEH1}. Moreover, these WSNs are and further will be participating in producing rapid inferences to support the increasingly complex tasks enabled by machine learning (ML) algorithms~\cite{netadapt,Origin}, often tweaked towards edge deployments, and applications of such edge-analytics is also exploding with the user and market demands. This represents a particularly challenging tension between energy availability and desired functionality, because the form factor constraints of the WSNs fundamentally limit active power, energy reserves, compute and communication capabilities.

For many WSNs, their participation in inference tasks has traditionally been limited to data collection and transmission, sometimes with modest preprocessing. While several studies have shown the benefits of performing more inference closer to the point of data collection~\cite{ResiRCA, intelligenceBeyondEdge,NVPMa, kang2017neurosurgeon, infocommDNNpart} and have applied these techniques to more powerful edge devices, their form-factor-imposed limited energy storage, low-power operation points, and deployment scenarios have been a major impediment in executing compute-intensive inference tasks directly on such platforms. In contrast, communicating the data, often after little preprocessing, although popular, is not cheap in terms of power requirement, and often poses a challenge for remotely deployed and ultra low power WSNs. Prior works, trying to tackle this conflict between computation, communication, power-requirement and quality of service (QoS), have pursued three major approaches: inference effort partitioning optimizations~\cite{kang2017neurosurgeon,batteryfree,infocommDNNpart,talhaDNN}, mitigation of energy provisioning limitations~\cite{IntBeyondEdge, chinchilla,NVPMa,ResiRCA,Origin}, and minimizing communication overheads~\cite{compression-kimura2005survey,quantcompression,compression-marcelloni2008simple,Ting-He-ArXiv, Ting-He-IEEE}.

One of the most emerging line of work aims to solve the energy provisioning problem at the edge by integrating energy harvesting (EH) to the sensor nodes while making them more capable performing complex compute intermittently, which has given rise to energy harvesting wireless sensor networks (EH-WSNs). Specifically, recent works~\cite{IntBeyondEdge, chinchilla} proposed EH, along with compiler/runtime optimizations and leveraging non-volatile processors (NVP)~\cite{NVPMa,ResiRCA}, to increase local compute at the edge. 
EH as a solution has been particularly interesting as a means to address the sustainability issue of battery backing trillions of future devices. More importantly, EH can help us build sustainable distributed sensing/monitoring infrastructure at virtually inaccessible places like oil-wells, mines, and even satellite orbits~\cite{orbitaledge,EHuse}. However, harvested energy is fickle in nature, and typically harvested sources only deliver scant microwatts of power (see Figure~\ref{Fig:EHsota} for an overview). The sporadic nature of harvested energy and the lossy nature of EH based storage and charging circuits calls for using the harvested energy directly to perform intermittent compute rather than storing energy for some distant future use. On this front, recent works~\cite{Origin,LuciaCheckpoint,chinchilla} have specifically optimized DNN inference execution at the EH-edge nodes by utilizing adaptive dynamic check pointing, intelligent scheduling and ensemble learning. Given the limitations of the EH budget, such approaches typically end up dropping many samples and not inferring from them locally. Importantly, they are often incapable of transmitting the raw data due to a lack of sufficient energy; for sensing tasks with modest inference requirements, performing inference and transmitting the result can take \textit{\textbf{less}} energy than transmitting raw data. However, to unleash the remote deployment, and sustainable, yet pervasive, computing capabilities WSNs, development of efficient \textit{energy harvesting WSNs} (EH-WSNs), both for sensing and edge-analytics, plays an essential role.

These wireless sensing (EH or otherwise) devices have long relied on compression techniques to mitigate data communication overheads~\cite{compression-kimura2005survey,quantcompression,compression-marcelloni2008simple}. However, when applied to low-dimensional sensor data,  classical lossy compression techniques tend to discard or distort some important features, which significantly degrades the inference accuracy. To mitigate the shortcomings of classical compression techniques, recent works~\cite{bachem2015coresets, Ting-He-ArXiv, Ting-He-IEEE} propose using \textit{coresets}, a data representation technique from computational geometry that preserves important, representative features when building a compressed form of the data, and thereby reducing the payload size while preserving data integrity for efficient edge communication. Although, with the help of coresets, one can efficiently offload minimal input representations to a more compute-capable device, performing accurate inference on coresets is non-trivial due to their low-dimensional nature.

From the aforementioned challenges, it is evident that we need a concoction of both hardware-driven and software optimized solutions to build next-generation EH-WSNs with the ability to perform fine-grained intermittent computing, while ensuring efficient network communication. Towards this, we propose \emph{Seeker}, a novel approach that leverages and extends coresets to efficiently execute DNN inference across a set of EH sensor nodes and a host mobile device. \textit{Seeker} focuses on building an efficient EH-WSN which can collaboratively work to maximize the inferences performed at the EH-edge nodes.  Furthermore, it then applies innovative coreset techniques to efficiently and intelligently offload unfinished compute tasks to a more capable host to further increase the inferences that can be performed.
Particularly, \textit{Seeker} augments its coreset formation with \textit{application-awareness} to form an energy aware, dynamically configured, and feature preserving payload with minimal communication footprint. \textit{Seeker} provides hardware acceleration support for coreset formation to make them computationally efficient, adaptive, and accuracy-preserving specifically for EH-WSNs. The following are the \textbf{primary contributions} of our work:

\squishlist
\item \textbf{Efficient Communication:~}We enable low data volume communication by developing extensions to traditional coresets that enhances their applicability to EH-WSN inference scenarios. Specifically, we introduce an \emph{activity-aware coreset construction} technique to dynamically adapt to both activity and the available harvested energy, while conserving maximum features of the data. This reduces the communication payload size by $8.9\times$. We also propose a \emph{recoverable coreset construction} technique, which helps reconstruct the original data from the compressed form with minimum (as low as $0.02\%$) accuracy loss. 
\item \textbf{Efficient Computation:~} We augment a state-of-the-art EH-sensor node with quantized DNNs to increase the number of accurate inferences at the edge (by up to $40\%$). We leverage data memoization to skip unnecessary compute saving inference execution time and energy.
\item \textbf{Efficient Hardware:~} We propose simple, low power, and low latency hardware to efficiently build coresets, further increasing the number of samples that can be inferred or transmitted under EH budget, and thereby significantly improving the accuracy over the state-of-the-art ($\approx5\%$). We develop a non-volatile hardware accelerator, with multiple quantization support, for efficient DNN inference. 

\item \textbf{Adaptability:~}Although \textit{Seeker} is meant for EH-WSNs, the coreset based data representation can easily be used in any commercial device for efficient communication.  

\item \textbf{Detailed Evaluation:~}We provide a detailed evaluation of our system and the proposed hardware design. Our evaluations show that, even when powered by an unreliable EH source, {\em Seeker}'s coreset-based optimizations result in better accuracy than that of a fully-powered system running a state-of-the-art classifier optimized for energy efficiency. Specifically, \emph{Seeker} reaches  86.8\% top-1 accuracy in comparison to the 81.2\% accuracy of the baseline system. 
\squishend

%% file: src/02BG-RW.tex
\begin{figure}
\centering
	\subfloat[ High-level overview of an energy harvesting system.]
	{
	\includegraphics[width=0.85\linewidth, height=2.5cm]{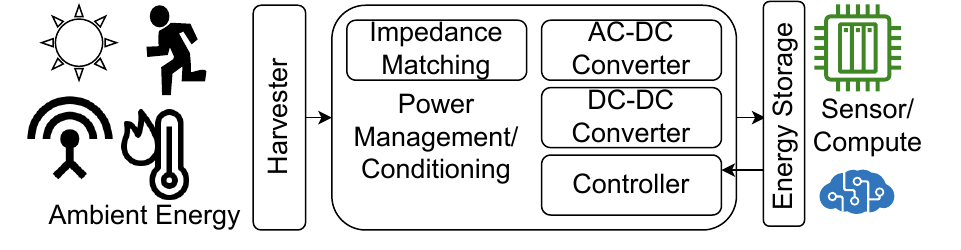}\label{Fig:EHPrimer}
	}
	\hfill \\
    \subfloat[Current state-of-the-art of EH-WSN.]
    {
    \includegraphics[width=0.65\linewidth]{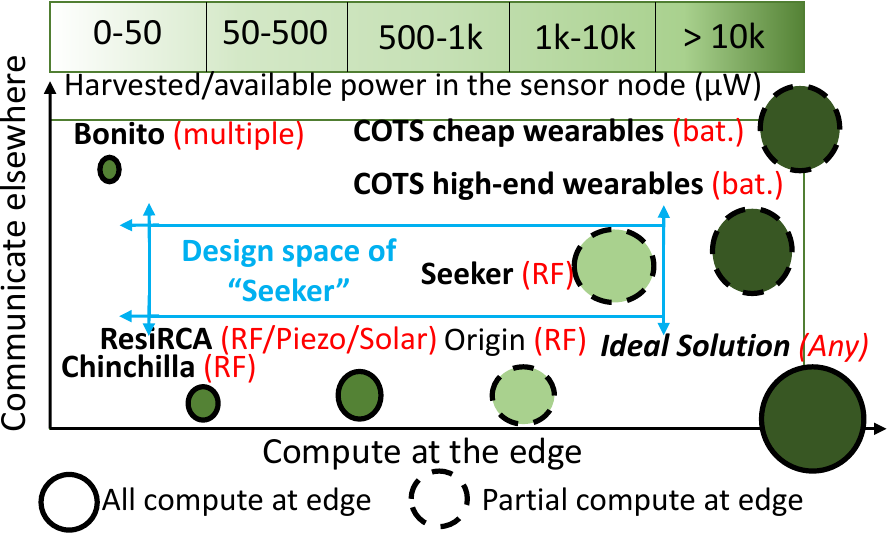}\label{Fig:EHsota}
    }
    \caption{A primer on energy harvesting systems: Figure~\ref{Fig:EHPrimer} shows the basic building blocks of an EH node equipped with sensing and computation. Some of the units change according to the harvested energy source. Figure~\ref{Fig:EHsota} shows the capabilities of the current SOTA. The size of the circle representing the solutions depicts the compute capabilities of the sensor nodes, the shade shows the available power, and their position on the axes approximates the amount of compute done on the node and the amount of reliability on external communication. The power source is denoted in \textcolor{red}{(Red)} (notations used in Figure~\ref{Fig:EHsota}: COTS: Commercial-off-the-shelf, Bat.: Battery, Bonito~\cite{batteryfree}, Chinchilla~\cite{chinchilla}, ResiRCA~\cite{ResiRCA}, Origin~\cite{Origin})}
    \label{fig:EHIntro}
\end{figure}

In this section, we provide a background of the current state-of-the-art in performing sensing and computations on EH-WSNs. We also describe the challenges in enabling complex compute on such devices and the need for hardware-software co-design to enable specialized intermittent computing in EH-WSNs. Finally, we define the scope of our work and focus on the problem specifics while alluding to probable solutions. 
Figure~\ref{Fig:EHPrimer} shows the basic building blocks of an energy harvesting sensing/computing unit. The harvested energy is typically stored in either an intermediate storage like a (super) capacitor~\cite{batteryfree}, or used for charging. For building scalable and sustainable infrastructure of battery-free EH-WSNs, the former is more feasible and will be our focus for this work. The fickle nature of harvested energy has posed a major challenge in performing any useful computation, as any useful forward progress gets lost when the traditional computing systems lose power. To tackle this, a significant amount of work has been done on check-pointing, and compiler level tweaks, which help maximize the forward progress on such devices~\cite{incidental, chinchilla, LuciaCheckpoint}. These solutions operate on augmented commercial off the shelf micro-controllers~\cite{TIEH}, or specialized products with traditional architecture~\cite{WISP}, and rely on their efficient prediction of power failure. While these software optimizations and judicious use of persistent storage works for smaller workloads like keyword spotting (e.g \textit{"Ok Google"} detection), they are inefficient for complex workloads (e.g. multi-sensor HAR, predictive maintenance etc.). 

These software-based solutions exhibit inefficiencies with respect to energy and time due to performing multiple save-and-restore cycles~\cite{intelligenceBeyondEdge,ResiRCA}: while some of these operations are necessary, unnecessary checkpoints will also be conservatively performed to ensure forward-progress. Therefore, recent works~\cite{NVPMa, incidental, NVPMicro, spendthrift, ResiRCA} propose the use of a NVP, where the non-volatility of the hardware itself takes care of saving and resuming the program execution. This reduces software overheads and latencies for handling power emergencies and hence can guarantee better QoS for complex and longer tasks even when power is deeply unreliable. Using an NVP and multiple harvested energy sources Qiu et al.~\cite{ResiRCA} demonstrates the possibility of performing complex DNN inference at the EH-Sensor itself. While an NVP ensures safe check-pointing for a given computation, current edge scenarios may require a device to be simultaneously performing multiple functionalities~\cite{iWatchBattery, AppleFall, AppleECG, Google-Assistant-Watch} and might be at energy scarcity. As a result, it is difficult to reliably run these complex tasks standalone on the edge device. Current devices adapt in one of three ways: \emph{1) Send all the sensor data to a connected host device, or cloud, to offload the compute and act only as a sensing and display device; 2) Process data on the device itself, potentially dropping or delaying tasks due to energy shortfalls; 3) A mix of the two models, where some computations do happen on the device while others are offloaded to balance compute, energy, and communication resources}; and typically, the latter is preferred, but it is non-trivial to find the optimal balance between what is to be done on the edge, what to be offloaded~\cite{kang2017neurosurgeon, taylor2018adaptive, zhao2018deepthings, eshratifar2018energy}, and \textit{how to efficiently offload}.

\noindent \textbf{\underline{Need for Specialized Hardware:~}}One of the major challenges in deploying learning tasks using EH-WSNs is to find the proper hardware platform. The current commercial-of-the-shelf (CotS) hardware capable of performing such compute are not energy efficient to run with all modalities of harvested energy since all of them do not have the same energy income (see Figure~\ref{Fig:EHsota}). For example, there has been significant work on enabling solar powered smart farming~\cite{smartfarmsolar, smartfarmsolar1}, but the same can not be done for smart manufacturing due to the lack of solar exposure and the low fidelity of the available EH sources such as vibration and RF (from WiFi or other sources). To estimate the required energy, we ran simple HAR inferences (optimized version of~\cite{HARHaChoi} for edge deployment using~\cite{netadapt}) on an Adafruit ItsyBitsy nRF52840 Express - Bluetooth LE~\cite{adafruitI} and found it to be consuming from 550mJ to 1.6J of energy (depending on the quantization). Compared to this, body movement and WiFi sources (the possible modalities of harvesting for HAR) harvests in order of milliwatts~\cite{batteryfree,ResiRCA}, making it almost impossible to have a feasible EH-WSN deployment, with the capabilities to perform modest learning tasks, using the CotS. Therefore, there has been a significant body of work~\cite{ResiRCA, Origin,NVPMa,NVPMicro} on developing appropriate next generation hardware (most of them on simulation). Although, we evaluate and show the communication cost savings of \textit{Seeker} on the battery backed CoTS hardware, we propose possible (simulated) hardware accelerator designs to fully deploying a EH-WSN capable of performing inference, compression and communication in harvested energy budget. 

\begin{figure}
\centering
	\subfloat[Completion with ERR]
	{
	\includegraphics[width=0.45\linewidth, height=2.5cm]{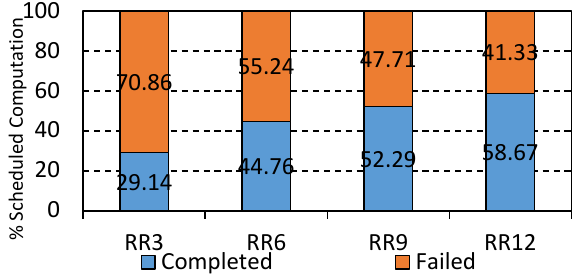}\label{Fig:frac-complete}
	}
	\hfill
    \subfloat[Accuracy of ERR]
    {
    \includegraphics[width=0.45\linewidth, height=2.5cm]{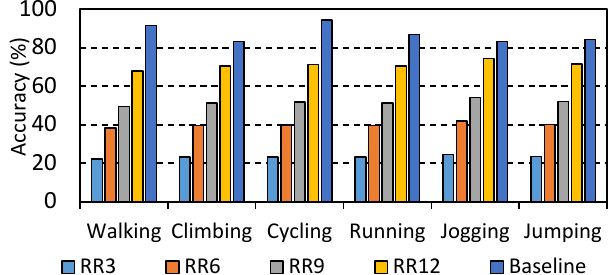}\label{Fig:RR-accuracy}
    }
    \hfill \\
    \subfloat[Accuracy vs quantizations]
    {
    \includegraphics[width=0.45\linewidth, height=2.5cm]{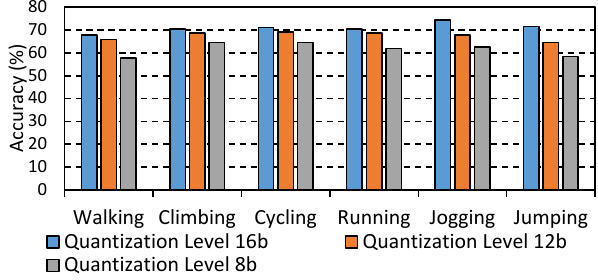}\label{Fig:quantized-accuracy}
    }
    \hfill
    \subfloat[Accuracy vs sub-sampling]
    {
    \includegraphics[width=0.45\linewidth, height=2.5cm]{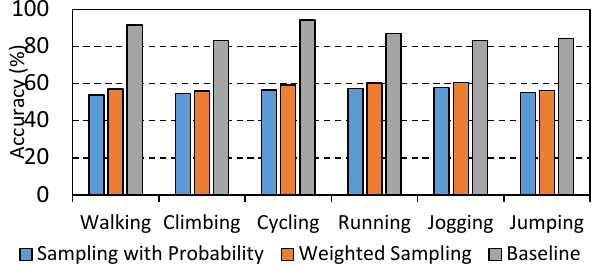}\label{Fig:subsampled-accuracy}
    }
    \caption{Accuracy comparison of various classical node-level optimization techniques. The Extended-Round-Robin policy (ERR)~\cite{Origin} takes a store-and-execute approach, and the number associated represents the ratio of store cycles vs execute cycles (e.g. RR3 is 3 store cycles followed by 1 execute cycle). The 'Baseline' model is a fully powered system with no energy restrictions, and the quantized model runs on harvested energy using a RR12 policy.}
    \label{fig:Projection}
\end{figure}

\noindent \textbf{\underline{Complex Compute on EH-WSNs:~}}To quantify the scope performing complex compute using EH-WSNs, we took human activity recognition (HAR) as a workload\footnote{Throughout the paper we evaluate many of our motivation results using HAR as a workload as it is one such application, where the (EH-)WSN, used as body area network, fits perfectly with RF or body movement as the harvesting source. HAR has the nuances of human introduced unpredictability and sensor induced noises. HAR has been pervasive enough given the rise of smart wearables and has been studied well enough to have ample access to resources to make a judicious evaluation. Further, we also evaluate one more emerging application from predictive maintenance domain.}, and performed experiments on the MHEALTH data-set~\cite{mHealth,mHealthDroid} (see Section~\ref{sec:evaluation} for data-set details) using the DNNs proposed in~\cite{HARHaChoi,HARHompel}, an energy harvesting friendly DNN hardware accelerator~\cite{ResiRCA} (to ensure that we are using the state of the art EH-WSN hardware) and recently proposed HAR-specific optimizations for EH systems~\cite{Origin}. Our analysis (see Figure~\ref{Fig:frac-complete}) shows that the state-of-the-art system still only finishes $\approx 58.7\%$ of the inferences scheduled on a sensor. Although accuracy can increase by further tuning duty-cycle, as shown in Figure~\ref{Fig:RR-accuracy}, the returns are diminishing, and indefinite increase of duty cycle is also not an option as that might lead to skipping important data to infer. We observe that the system used in~\cite{Origin} does not aggressively employ quantization, which is a commonly used technique~\cite{drq-isca2020} to reduce both compute and transmission energy in DNN tasks. Our analysis, as shown in Figure~\ref{Fig:quantized-accuracy}, shows accuracy as a function of quantization (we took the approach of performing post training quantization and fine-tuned the DNN to work with reduced bit precision instead of training the DNN from scratch with a reduced precision). The quantized DNNs benefit from lower compute and memory footprints, but need specialized fine-tuning and often suffer from lower accuracy. Similarly, other  approximation-via-data-reduction techniques, such as sub-sampling, did not perform inference with a desirable accuracy. \textit{Collectively, the aforementioned figures demonstrate that the harvested energy budget is insufficient to perform {\em all} inferences with acceptable accuracy on currently proposed EH-WSN systems.} Therefore, to complete all the scheduled computations, and thereby to improve accuracy, the system must rely on another device (e.g. a mobile phone), where sufficient resources are available to complete any remaining inference, \textit{if the data can be sent from the sensor}. Said coordinating device completes the rest of the computations and finally, aggregates them with the ones completed in the sensor nodes. The challenge here is to \textit{send the data efficiently}, since communication is an expensive task and especially challenging for EH-WSNs~\cite{batteryfree} thanks to their fickle and ultra-low energy budget. The obvious solution is to reduce the communication data volume by compressing the data before transmitting. This also reduces energy footprint and the probability of data packet loss.

\begin{table}
\centering
\begin{adjustbox}{max width=\linewidth}
\begin{tabular}{|c|c|c|}
\hline
\textbf{Algorithm}    & \textbf{Compression Ratio} & \textbf{Accuracy Loss (\%)} \\ \hline
Fourier Decomposition & 3 - 5                      & 9.1 - 18.3                  \\ \hline
DCT                  & 3 - 5                      & 5.8 - 16.2                  \\ \hline
DWT                 & 3 - 6                      & 5.3 - 12.7                 \\ \hline
Coreset               & 3 - 10                     & 0.02 - 0.76                 \\ \hline
\end{tabular}
\end{adjustbox}
\caption{Accuracy trade-off of different compression techniques: Low-dimensional data loses important features under lossy compression, dropping inference accuracy significantly compared to the original data. Details on Coreset are available on Section~\ref{sec:4}. (Notations used: DCT: Discrete Cosine Transform, DWT: Discrete Wavelet Transform.}
\label{tab:compression-table}
\end{table}
 
\noindent \textbf{\underline{Challenges with Data Compression:}~} Using standard compression algorithms, like discrete cosine transform, discrete wavelet transform, and Fourier decomposition etc., to minimize the communication overhead is not a viable solution~\cite{compression-marcelloni2008simple}. This is partly because we need a very high compression ratio with very low power. Secondly, these compression algorithms are not context-aware, and hence lose relevant features during the process of compression resulting in degraded inference accuracy (refer Table~\ref{tab:compression-table} for details). A key insight is that, while these compression techniques work very well for high dimensional data (e.g. images), inference on low-dimensional sensor data (such as inertial measurement unit or IMU vibration data) is much more sensitive to lossy compression as separating between features might be difficult to do. And we will not achieve a sufficient compression ratio from lossless approaches either. Therefore, the standard data compression techniques are not very useful, let alone their energy efficient (such as quantized versions~\cite{quantcompression}) counterparts. For data compression in EH-WSNs, we need the compression algorithm to be \circled{1} \textbf{light weight} (for energy efficiency), \circled{2} \textbf{feature preserving} (for higher accuracy), \circled{3} \textbf{having a high compression ratio} (for communication efficiency), and \circled{4} \textbf{context agnostic} (for better generalization); i.e., our deployment scenario demands a \textit{smaller representative form} of the data that still \textit{preserves enough  application-specific features to perform meaningful classifications in a given DNN}.

\noindent \textbf{\underline{Why Coresets?}}
The aforementioned requirements motivate us to consider \emph{coresets}  for forming representations of the original data. Coresets, primarily used in computational geometry~\cite{bachem2015coresets},  have been recently used~\cite{Ting-He-ArXiv, Ting-He-IEEE} for machine learning and sensor networks. Since coresets were designed to preserve the geometry of the data, we believe that they can be crafted to preserve features, and therefore be useful for performing accurate inference in subsequent stages, and thereby satisfying \circled{2}. Furthermore, constructing coresets do not need any application information, i.e. they are application/data agnostic and can represent any form of data (IMU~\cite{Ting-He-ArXiv}, Image~\cite{6907021}, DNN feature map~\cite{coresetDNN, coresets-compression-ECCV,coresets-traning-GMM,coresets-traning-pmlr}). This fulfils requirement \circled{4}. They are also an effective way to construct a representation of the data set with high compression ratios~\cite{practicalcoresets,coresets-compression-ECCV} without incurring unacceptable accuracy losses and thus useful for achieving \circled{3}. For the DNNs in question, coresets can achieve sufficient compression ratios to make communication energy-competitive with computation, as well as opening up new opportunities for optimizing DNN inference on the coreset, rather than original data.  Finally, most of the coreset construction algorithms are simple (hence can achieve \circled{1}) and do not need complex operations (like cosine, exponential, etc.~\cite{bachem2015coresets,practicalcoresets,Ting-He-ArXiv,Ting-He-IEEE}, they can also be quantized~\cite{Ting-He-IEEE} to further reduce their computation and memory footprints. Motivated by this, we explore possibilities of designing an efficient synergistic sensor-host ecosystem (involving the EH-WSNs and host), where we try to maximize the compute at the sensor nodes, yet for the incomplete tasks, we use coresets to compress and send the data to the host where the rest of the computations could occur.

%% file: src/03DesignSpaceExp.tex
\begin{figure}
\centering
\includegraphics[width=0.75\linewidth]{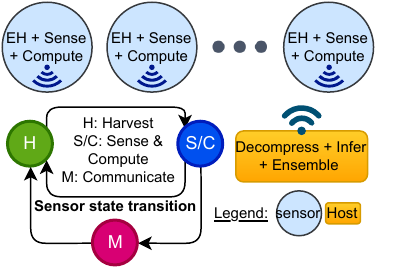}
  \caption{An example of EH Sensor-Host ecosystem - the sensor transitions between multiple states and executes the compute as store and execute fashion~\cite{Origin}. The host receives the data in compressed form for the unfinished portion, decompresses it, runs inference and finally ensembles the results from multiple sensors to improve accuracy and robustness.}
  \label{Fig:ecosystem}
\end{figure}
Since data communication in a sensor host ecosystem (Figure~\ref{Fig:ecosystem}) consumes substantial power, we rely on coresets as an efficient way to lossily communicate the features with minimal information degradation. The coreset construction techniques need to be extremely lightweight while preserving key features to justify the computation-communication trade-offs in energy and latency. To this end, we explore two different kinds of coreset construction techniques. 

\subsection{Coreset Construction Techniques}
\begin{figure}
\centering 
\includegraphics[width=0.65\linewidth]{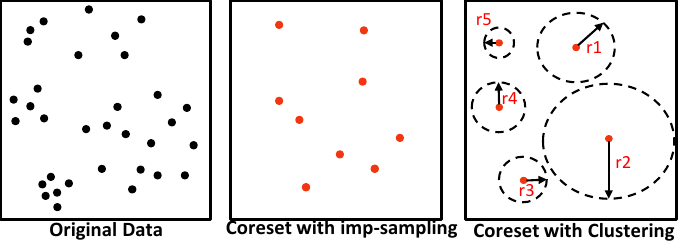}
\caption{A toy example of the coreset construction techniques in \emph{Seeker}. Imp-sampling uses a probability based  importance sampling; clustering preserves the geometric shape of the original data. In each case, the points/values in \textcolor{red}{red} are communicated to the host.}
\label{Fig:Coresets}
\end{figure}

\textbf{\underline{Coreset Construction Using Importance Sampling:}~}An easy way to build a representation from a data 
distribution is to perform importance sampling~\cite{practicalcoresets, bachem2015coresets}, i.e. give more importance in choosing the data which are unique and, in our case, contribute significant to the inference (i.e. having a high enough magnitude in the frequency response of the sensor signal). The intuition is that any importance
sampling scheme produces an unbiased estimator~\cite{practicalcoresets}. To preserve the temporal and frequency features, we ensure sampling data which are far enough from each other to build a better representation. The entire process of importance sampling uses simple arithmetic operations and is therefore viable in energy-scarce situations.
The host can take the sub-sampled data and perform inference. The caveat is to have a model trained on the sub-sampled data, which can be done as an one-time step. Although the sub-sampling might lead to poor inference accuracy, in our experiments, with iso-compression ratio, importance sampling based coresets still outperforms classical compression techniques. Figure~\ref{Fig:Coresets} shows a toy example of importance sampling in a 2D data set. Observe that the selected points (in \textcolor{red}{red}) are approximating the original distribution.

\noindent \textbf{\underline{Coreset Construction Using Clustering:}~} Although importance sapling based coreset construction is computationally inexpensive, it suffers from accuracy loss because it doesn't explicitly preserve the intricate structure of the data points. To address this, we also utilize coreset construction using k-means clustering~\cite{Ting-He-ArXiv, Ting-He-IEEE, practicalcoresets}, which separates the data points into a set of k (or fewer) N-spherical clusters and represents the geometric shape of the data by using the cluster centers and cluster radii (Fig.~\ref{Fig:Coresets}). These are then communicated to the host device for inference. Since clustering better preserves the geometry of the distribution, we observe that inferences with coresets constructed using clustering are more accurate than using importance sampling, and therefore can be preferred over the former whenever there is enough energy.

\vspace{-8pt}

\subsection{Communication vs Accuracy}
\begin{figure}
  \centering
  \includegraphics[width=0.8\linewidth]{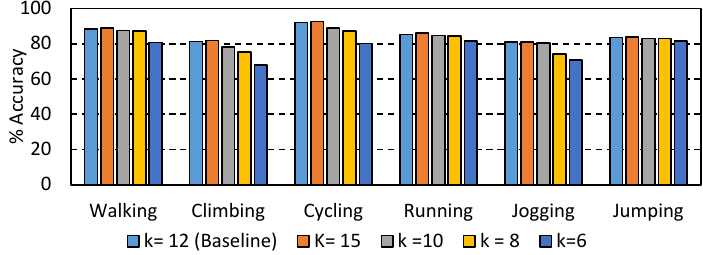}
  \caption{Accuracy with different \#clusters (k).}
  \label{Fig:AAC-Accuracy}
\end{figure}

We can tune the aforementioned coreset construction techniques allow a variable number of features depending on the available energy, i.e. for importance sampling, we can limit the number of points to choose, and similarly, for clustering we can limit both the number of clusters and the number of iterations. 
However, even after preserving important features, the constructed corests are  lossy representation of the original data. Therefore, when performing inference on the compressed coresets representation, the inference accuracy goes down, albeit not significant compared to other lossy compression methods (we can again refer to Table~\ref{tab:compression-table} for the relevant comparisons). 
This leaves an optimization space in trading between communication cost vs. accuracy, i.e. \textit{whether to construct strict and low-volume coresets and lose accuracy or to preserve maximum data points and pay for the communication cost}. We perform an analysis on the MHELATH~\cite{mHealth, mHealthDroid} data set (we take a overlapping moving window of 60 data points sampled at 50Hz from 3 different IMUs, overlap size: 30 data points) to find a trade-off between the coreset size (directly related to the communication cost) and the inference accuracy. Empirically, we observe that accurately preserving the features for each class requires \textbf{20 data points} using importance sampling or \textbf{12 clusters} (see Fig.~\ref{Fig:AAC-Accuracy}) using clustering based techniques. Going above 12 clusters did not significantly improve accuracy. This further motivates us to look for opportunities in the data distribution to improve the compression ratio.

As the DNN models were designed to infer on the full data, we retrain the DNN models to recognize the compressed representation of the data and infer directly from that (both from the importance sampling and clustering). As the coreset formation algorithms are fairly simple~\cite{bachem2015coresets, practicalcoresets, Ting-He-ArXiv, Ting-He-IEEE}, it does not take much latency or energy to convert the raw sensor data into the coreset form even while using a commercial-off-the-shelf micro-controller (like TI MSP430FR5969~\cite{TIEH}). This allows the EH-sensor to opt for coreset formation followed by data communication to the host device as an energy-viable alternative to local DNN inference on the original data. In our example case, transmitting the raw data (60 data points, 32bit floating point data type) needs \textbf{240 Byte}s of data transfer, and with coreset construction and quantization we can limit it to \textbf{36 Bytes} (for 12 clusters, each cluster center is represented by 2 Bytes of data, and radius represented by 1 Byte data), thereby \textit{reducing the data communication volume by 85\%}. The host runs inference on the compressed data to detect the activity (with an accuracy of 76\%). However, due to this reduced accuracy, the sensor only takes this option iff it does not have enough energy to perform the inference at the edge device (either in the 16bit or 12bit variant of the DNN - more details on DNN design is presented in Section~\ref{sec:4}). This raises a question: \textit{is it possible to generate a more \textbf{useful} approximation, via reconstruction, of the data that we lost while forming the coresets?} This problem has not been explored in details, as coresets are typically considered as an $\alpha-$approximate representation of the data ($\alpha$ being the error/approximation parameter)~\cite{bachem2015coresets} and never needed proper recovery. However, thanks to  the low dimensional nature of many sensor data, reconstruction of original data from coresets becomes an essential step.

\subsubsection{Data Memoization:}
Given our focus on ultra low power energy harvesting devices, any opportunities to reduce computation and communication can noticeably augment the performance and efficiency of the entire system. We look into data memoization as one such opportunity. For two instances of the same class, there should be a very high correlation in the sensor data. We empirically measure this by testing for correlation between the sensor signatures of different classes. Conservatively, we choose a correlation coefficient $\geq 0.95$ to predict that the two activities are the same, and hence skip the inference altogether and just communicate only the results to the host. We store ground truth sensor data pattern for all possible labels, and when new data arrives, we find the correlation of the sampled data against the ground truth data, and if any of the correlation coefficient comes out to be $\geq 0.95$, we choose to ignore further inference computation and only communicate the classification result to the host for further processing. Note that choosing the correlation threshold entirely depends on the application and user preference.
 
 \begin{figure}
  \centering
  \includegraphics[clip,width=0.65\linewidth]{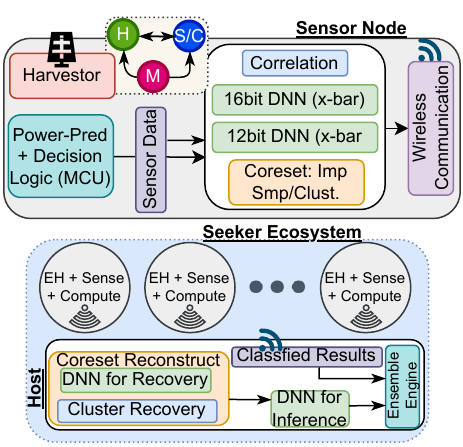}
  \caption{Overall system design of Seeker}
  \label{Fig:system_design}
 \end{figure}

\subsubsection{Recoverable Coreset Construction:}\label{sec:recover} The primary reason the accuracy of inferring on coreset data is lower than that of the original model is the loss of features. Typically, the sensor data are low dimensional, and hence even with a good quality of coreset construction, it is difficult to preserve all the features. However, while inferring at the host, if we are able to recover the data or reconstruct it with minimum error, the accuracy can easily be increased. 

\begin{figure} 
 \centering
    \subfloat[Recovering a cluster with uniform random re-distribution.]
    {%
     \includegraphics[clip,width=0.75\linewidth]{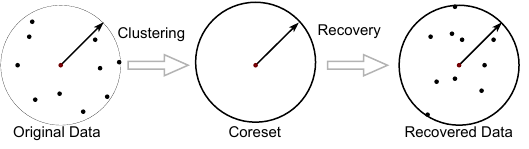}%
    \label{Fig:clusterRecovery}
    }
    \hfill \\
    \subfloat[Recovering a sub-sampling with GAN.]
    {%
     \includegraphics[clip,width=0.75\linewidth]{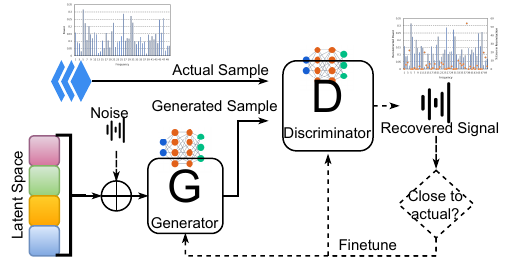}%
    \label{Fig:DPRecovery}
    }
    \hfill
    \caption{Recovering data from the coresets.}
    \label{fig:AllRecovery}  
\end{figure}

\noindent \textbf{\underline{Clustering Coreset Recovery:}} Clustering preserves the geometry of the original data by representing them as a set of N-spherical clusters represented with a center and a radius. In the process of coreset construction we only preserve the coordinates of the centers and the radii of the clusters, and hence miss the coordinates of the points inside the clusters. However, any random distribution of the lost points in the cluster could provide us with a $2r-$approximate representation of the original distribution (where $r$ is the radius of the cluster; refer Figure~\ref{Fig:clusterRecovery} for a toy example). However, to achieve this, we need some extra information about the clusters. The standard method of clustering-based coreset construction keeps the cluster center and cluster radius, which gives the geometrical shape of the entire data. Extending this with \textbf{the point count for each cluster} allows for reconstruction of data in the original form that can be processed by DNNs trained on full-size data. These reconstructed data sets can be synthesized simply by uniformly distributing the points within each cluster. Although the intra-cluster data distribution will be different from the original, it will still preserve the overall geometry with a certain degree of approximation which the DNN could learn to accommodate.

Experimentally, on the MHELATH dataset, we observe that inferring on the synthesized reconstructions of cluster based coresets can achieve an accuracy of $\approx 85\%$. The reconstruction feature at the host comes with little to no overhead for the host (given the host devices have considerably more compute than the sensor nodes). The addition of the recovery parameter (number of points per cluster) needs \textbf{4 more bits} (in our experiments, we never observe any clusters having more than 16 data points) of data per cluster, bringing the total data communication volume to \textbf{42 Bytes}, which is still a significant $5.7\times$ less in comparison to the original 240 Bytes needed to communicate the raw data in our setup. However, since clustering based coreset construction is more expensive than the importance sampling based coreset construction, it is not always possible to build a recoverable coreset at the edge, unless we figure out a to recover the lost points while we perform importance sampling.

\noindent \textbf{\underline{Importance Sampling Coreset Recovery:}} Unlike clustering, when we construct a coreset with importance sampling, we typically have no information regarding the lost data points. We hypothesize that the dropped sample should contain, although not important, sensor specific artifacts. And these artifact must have some pattern, if modeled correctly, could represent the lost data. Towards this, we designed and trained a generative adversarial network (GAN, see Figure~\ref{Fig:DPRecovery} for the structural details) to recover the lost samples of the importance sampling. As training parameters, we provide some statistical parameters (specifically mean and variance) of the signal and random noise to the generator, and the generator generates the lost signals. The discriminator tries to discriminate between the actual data and the synthesized data. We fine-tune the network until the discriminator is fooled sufficiently to distinguish between the original data and the recovered data.   
Considering the fact that we do have access to the sensor data to train the learning algorithm, we can use the same data to train the GAN and with sufficient data, the discriminator could generate the lost signal with minimum error. Our experiments show that the deviation from the original signal, in most cases, is limited to $\le15\%$. However, in some pathological cases, the error at times goes close to $60\%$, and we believe them to be generated artifacts which are common side effects of the GANs\cite{ganerror}.  Our experiments suggests that inference on the GAN recovered signal is almost as good as (about $2\%-4\%$ difference in accuracy) the inference on the recovered cluster signal. The recovery policy can be implemented as a simple generator network in the host. Although, the training of the GAN is complex and involves multiple networks as well as hyper-parameters tuning, the generator network itself is very small (\textit{few hundred thousands} of parameters depending on the sensor data). 

\vspace{-8pt}

%% file: src/04Seeker.tex
By leveraging the coreset construction techniques discussed in Section~\ref{sec:DSE}, we design \emph{Seeker: A synergistic sensor host ecosystem}. Figure~\ref{Fig:system_design} gives a pictorial representation of the overall design of \emph{Seeker} and its various components.
\textit{Seeker} leverages the concept of NVP, and employs a flexible store and execute method using the state of the art ReRAM crossbar architecture~\cite{Origin} to perform inference at the edge. It augments the sensor nodes with two different quantized DNNs (16 bit and 12 bit) to  increase the number of completed inferences at the sensor node itself. Prior studies~\cite{netadapt,ResiRCA,drq-isca2020} and our empirical analysis on the quantization vs accuracy trade-offs (see Fig.~\ref{Fig:quantized-accuracy}) indicate the 16 and 12bit precision to maximize the accuracy of the inferences while minimizing the energy consumption. Moreover, we also implement the memoization option so that it does not have to repeat inferences if it encounters similar data, thereby saving substantial energy as well as delivering results with extremely low latency. However, even with all these optimizations, due to the fickle nature of EH, \textit{Seeker} cannot finish all the inferences at the edge and must communicate with a host device. To minimize the data communication overhead between the sensor-node and the host device, \textit{Seeker} utilizes coresets to build representative, yet compressed, forms of the data. 

To cater towards the fickle EH budget, we use the two different coreset construction techniques, described in Section~\ref{sec:DSE}: a cheaper, less accurate formation (importance sampling) and a more expensive, yet accurate formation (K-means). Transmitting coresets rather than raw data greatly improves the energy efficiency of communication to the host, when required, and effectively increases the number of completed inferences, thereby increasing overall accuracy. Depending on the incoming data and the EH budget, the sensor decides whether to skip compute, perform an inference at the edge, or form a coreset to offload the inference to the host. The host, after obtaining information from multiple sensors, performs any further required computation and uses ensemble learning~\cite{Origin} to give an accurate classification result. Note that, unlike prior EH-WSN systems~\cite{Origin}, the role of the host device here is not limited to just result aggregation; rather, the host participates and performs inference when the sensors do not have enough energy and choose to communicate the data (in the form of coresets) to the host. In this section, we will explain, in detail, the overall execution workflow of the \emph{Seeker} system, followed by the the detailed design of the hardware support to maximize its energy efficiency.

\subsection{Decision Flow: From Sensors to the Host}
\begin{figure}
  \centering 
  \includegraphics[width=0.65\linewidth]{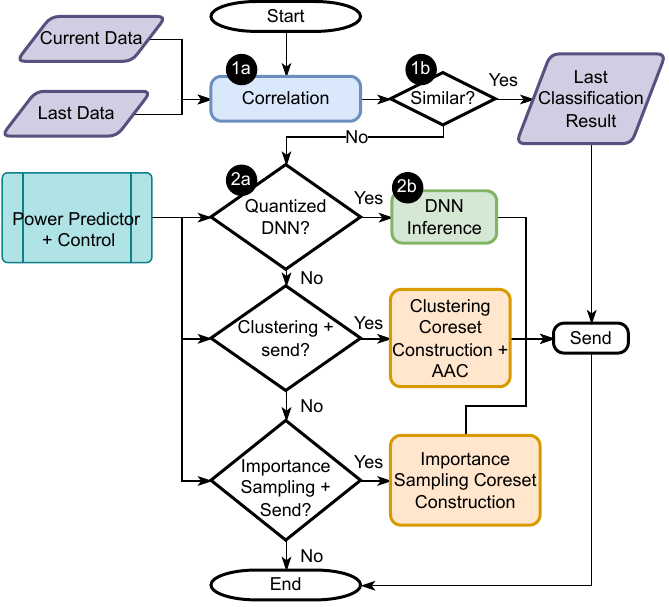}
  \caption{Decision flow of Seeker.}
  \label{Fig:DecsionFlowSeeker}
\end{figure}

Figure~\ref{Fig:DecsionFlowSeeker} depicts a flow chat showing the decision process taken in the sensor nodes to navigate between each components. Each sensor has a data buffer that collects the data points for classification (implemented using a 60 $\times 3$ FIFO structure of 4Byte cells to store the floating point data. The $\times 3$ caters towards the multiple channels of the sensor. The moving window is designed using a counter to shift the streaming data.) The sensor also stores one ground truth trace for each activity. The sensor computes the correlation (\circled{1a}) between the stored ground truth and the current data. If the correlation coefficient is $\geq threshold$ (\circled{1b}) the sensor skips all computation and sends the result to the host. Otherwise, the sensor prioritizes local computation and, with the help of a moving average power predictor~\cite{Origin}, predicts whether it can finish the quantized DNN inference with the combination of stored energy and expected income (\circled{2a} and \circled{2b}).  If energy is insufficient for DNN inference, the sensor will use coreset formation to communicate the important features to the host, which completes the inference. Since the clustering based coreset is typically more accurate then those formed by importance sampling, the former is preferred, when possible. We increase the frequency of cluster-based formation by using custom, energy efficient hardware. With the help of an activity-aware and recoverable coreset construction and low-power hardware design, we can efficiently communicate inferences or compressed data to the host device with minimum power and latency overheads. 

\emph{Seeker}, accounting for the available energy budget, considers the following decisions:
\textbf{D0:} Test for data similarity using correlation, and if similarity is found then communicate the results to the host;
\textbf{D1:} DNN at sensor with raw data + Communicate the results to the host;
\textbf{D2:} Try Quantized DNN inference and communicate the results to the host;
\textbf{D3:} Clustering based coreset construction at the sensor, and communicate the coreset to the host; host runs DNN inference on the reconstructed data; and
\textbf{D4:} Importance sampling based coreset construction at the sensor and communicate the coresets to the host; host recovers the original data with the pre-programmed generator, and performs inference with the recovered data.  
Table~\ref{tab:energy-numbers-table} lists the energy requirements of each of these decisions.

\begin{table}
\centering
\resizebox{0.45\textwidth}{!}{%
\begin{tabular}{|l|l|l|l|l|}
\hline
\textbf{Strategy} & \textbf{Sensor Energy} & \textbf{Comm. Energy} & \textbf{Total Energy} & \textbf{Avg Acc (\%)} \\ \hline
D0 & 0.54  & 8.27  & 8.81  & --\\ \hline
D1 & 29.23 & 8.27  & 37.5  & 80.03\\ \hline
D2 & 16.58 & 8.27  & 24.85 & 77.37\\ \hline
D3 & 1.07  & 15.97 & 17.04 & 78.30\\ \hline
D4 & 0.87  & 15.97 & 16.84 & 85.30\\ \hline
raw data & -- & 70.16 & 70.16 & 87.23\\ \hline

\end{tabular}%
}
\caption{Energy breakdown of different Seeker strategies (in $\mu$Joules). The accuracy reported is the average case over $\ge 1000$ inferences.}
\label{tab:energy-numbers-table}
\end{table}

\subsection{Efficient Hardware Accelerator}
\label{hardwaresupport}
Energy harvesting brings challenges in both average power levels and power variability. Performing DNN inference under such conditions often limits exploitation of inherent DNN parallelism within the energy budget. Therefore, many prior works use custon DNN accelerators, typically based on (non-volatile) resistive RAM (Re-RAM) based~\cite{ResiRCA, Origin} crossbar architecture, to perform DNN inference on EH-sensor nodes. \textit{Seeker}'s inference engine follows the design proposed in ResiRCA~\cite{ResiRCA} and modifies it to cater towards new quantization requirements: We have two different instances of the Re-RAM crossbar in our system - one for the 16bit model and one for the 12-bit model. The nonvolatile nature of the Re-RAMs helps in performing intermittent computing with the harvested energy. Moreover, techniques like loop tiling and partial sums~\cite{ResiRCA} can further break down the computation to maximize forward progress with minimum granularity. 

\begin{figure}
  \centering 
  \includegraphics[width=\linewidth]{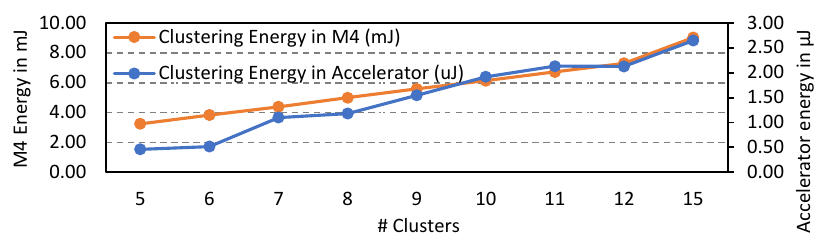}
  \caption{Energy consumption comparison of the clustering accelerator vs the SoTA hardware}
  \label{Fig:accvssota-cluster}
\end{figure}

While DNN inference is already accelerated in the sensor node, the addition of a general-purpose processor for coreset computation would be energy inefficient. Specifically, the computation requirement is fixed and does not require the overheads to support generality. Therefore, we add a low power, low latency coreset construction hardware. Comparing the energy consumption of the proposed clustering accelerator with  Adafruit ItsyBitsy nRF52840 Express - Bluetooth LE~\cite{adafruitI} suggests the accelerator to be $\approx 3.7\times 10^{3}$ times energy efficient (refer Figure~\ref{Fig:accvssota-cluster}).
Both the coreset construction algorithms follow a sequence of multiply and add/subtract operations followed by averaging, and hence can be simply designed with few logic units. Moreover, as we are operating with lower data volume, these operations can be parallel (for example, the clustering hardware simultaneously works on all the cluster formations). The bigger challenge is posed by the requirement of a variable number of iterations for these algorithms to converge, the number of clusters/samples required, etc.  To efficiently design the hardware and configure its parameters, we run several experiments and empirically arrive at the following conclusions: \textbf{1.} The clustering finishes within 4 iterations and, for importance sampling, it takes up to 7 iterations. \textbf{2.} None of the clusters have more than 16 points during any clustering. \textbf{3.} We need not store all the points in either cases at every iteration, rather the clustering hardware needs to store the sum, the radii of the clusters, the number of points per cluster, and the importance sampling hardware needs just the points. \textbf{4.} Storing the radii helps in easily selecting the points in the subsequent iterations. 

\vspace{-8pt}

%% file: src/05Evaluation.tex
In this section, we describe our methodology to evaluate \textit{Seeker}. We start with implementing \textit{Seeker} using the CotS  Adafruit ItsyBitsy nRF52840 Express~\cite{adafruitI} as the sensor compute node and a Google Pixel 6 Pro as the host node. Further, we describe how the design performs in simulated state-of-the-art hardware accelerator specifically designed for EH purpose. We look into two different applications: multi sensor human activity recognition (HAR), and bearing fault detection for predictive maintenance and compare the results with multiple baselines designed for ultra-low-power as well as EH environments. 
\begin{figure}
  \centering 
  \includegraphics[width=\linewidth]{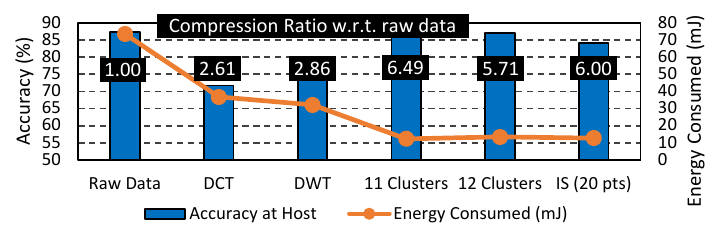}
  \caption{\textit{Seeker} vs other compression techniques}
  \label{Fig:seekercots}
\end{figure}
\subsection{\textit{Seeker} on Commercial Hardware}
Although Seeker is designed for EH-WSNs, the efficient communication mechanism for low dimensional sensor data can still be useful for the current commercial devices. 
Most of the ultra-low-power ($\le 30mW$) micro-controllers are not equipped with complex multiply-accumulate units to efficiently perform DNN computations, and hence are suited to collect, compress and send the data to a host (Pixel 6 Pro) and then the host  decompresses (or recovers) the data and performs the inference. We used the inertial measurement unit data of MHEALTH~\cite{mHealth} dataset as the sensor data, which is pre-processed and compressed at the Adafruit compute node and then sent over Bluetooth low-energy to the host. We used Circuit Python~\cite{CircuitPy} and Mu Editor~\cite{MuE} to implement the compression and the communication algorithms in the Adafruit board, and TensorFlow lite~\cite{tensorflow} to deploy the DNN inference at the host. We evaluate the efficiency, both in terms of compression ratio, energy consumption and accuracy preservation, of the recoverable clustering and recoverable importance sampling algorithms against three other popular methods: 1) sending raw data without compression; 2) compression using DCT; 3) Compression using DWT. We measure the energy consumption and inference accuracy over 1000 iterations to provide an average fair estimate. As depicted in Figure~\ref{Fig:seekercots}, \textit{Seeker} out performs both DCT and DWT in compression ratio, and the recovery feature of \textit{Seeker} helps preserving inference accuracy close to the original raw data.

\subsection{\textit{Seeker} for Activity Recognition}

 \begin{figure*} 
    \subfloat[Data volume with dynamic coresets]
    {%
     \includegraphics[clip,width=0.3\linewidth, height=3cm]{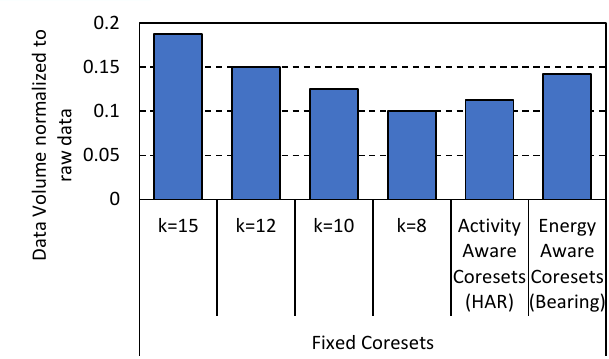}%
    \label{Fig:clusters-datavol}
    }
    \hfill
    \subfloat[Fraction of inferences completed with different EH sources]
    {%
     \includegraphics[clip,width=0.3\linewidth, height=3cm]{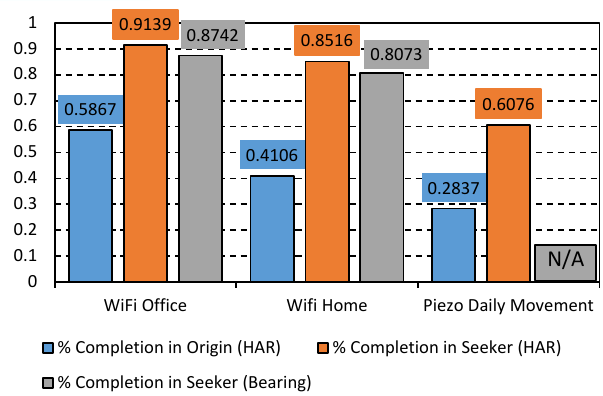}%
    \label{Fig:Sensitivity-fracCompleted}
    }
    \hfill
    \subfloat[Distribution of compute off-load to different components]
    {%
  \includegraphics[clip,width=0.3\linewidth, height=3cm]{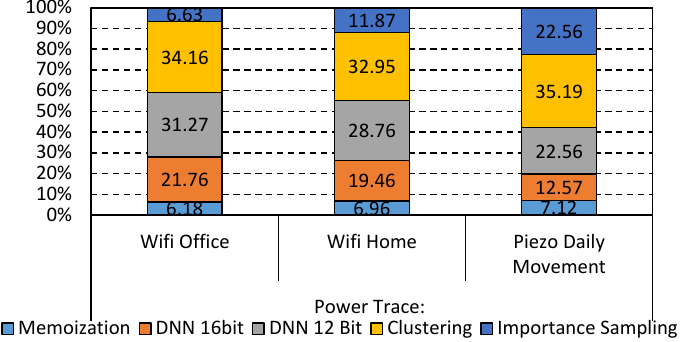}%
  \label{Fig:Sensitivity-fracOffloaded}
    }
    \caption{Accuracy and communication efficiency of \emph{Seeker} with different data sets and its sensitivity towards various EH sources.}
    \label{fig:AllSensitivity}  
\end{figure*}

\begin{figure*}[ht]
\centering
    \subfloat[Accuracy with MHEALTH dataset]
    {
    \includegraphics[width=0.48\linewidth
    , height=2.5cm
    ]{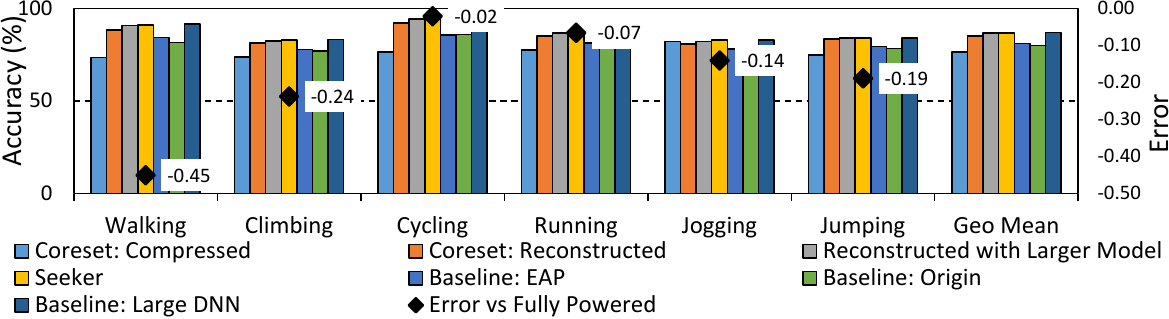}\label{Fig:MHEALTH-accuracy}
    }
    \hfill
    \subfloat[Accuracy with PAMAP2 dataset]
    {
    \includegraphics[width=0.48\linewidth
    , height=2.5cm
    ]{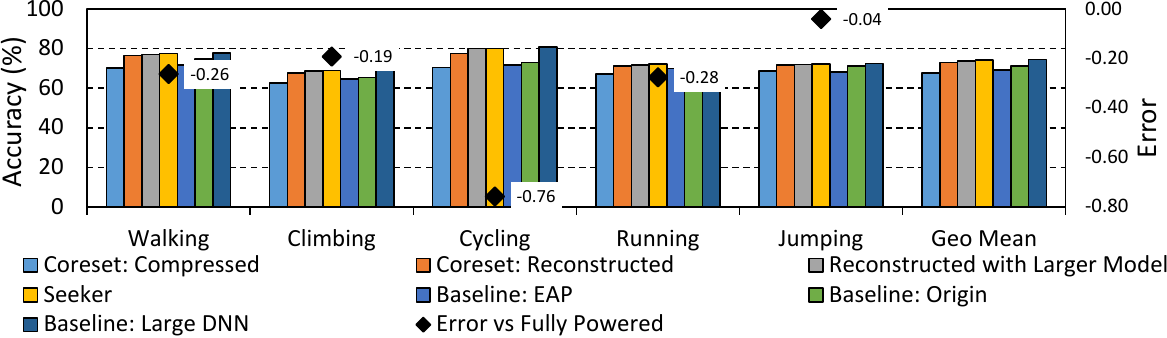}\label{Fig:PAMAP-accuracy}
    }
\caption{Accuracy and communication efficiency of \emph{Seeker} with different data sets and sensitivity study.}
    \label{fig:AllAccuracy}  
\end{figure*}
Human Activity Recognition (HAR) using body area network is becoming mainstream on most of the warble devices. Moreover, the pervasive nature of HAR along with ample opportunities to harvest energy, makes HAR on body area network quite interesting. Therefore, as a case study, we simulate an entirely EH body area network using all the components of \textit{Seeker}; specifically, to leverage intermittent computing using EH only, we simulate HAR on the hardware described in Section~\ref{hardwaresupport}. This includes three different sensors located at left ankle, right arm, and chest. Each sensor has (i) sensing element a.k.a Inertial Measurement Unit that collects acceleration data), (ii) two DNN Re-RAM crossbar (16bit \& 12bit) built using XB-SIM~\cite{XBSIM}, (iii) two coreset computation engines synthesized using Design Compiler~\cite{SynDC}, (iv) an energy harvester unit which is modeled after real-world energy harvester trace data obtained from the works by Qiu et al.\cite{ResiRCA} and Geissdoerfer et al.~\cite{batteryfree} (the specifics of the energy-harvesting mechanism producing the power trace are beyond the scope of this work.t), (v) a simple moving average power predictor power predictor and, (vi) low energy communication unit which uses IEEE 802.15.6. We model the communication energy based on the current state-of-the-art low energy communication systems~\cite{ulpComm1, batteryfree}. We utilize a simulation driven approach as multiple components, including the crossbar, coreset engine etc., are specialized hardware that are not commercially available. System development in the ultra-low-power space fundamentally spans the device technology, microarchitecture, architecture, and networking fields, and understanding the design space of next-generation EH-WSNs requires incorporating proposed advances from all areas. The crossbar simulator~\cite{XBSIM, ResiRCA} accurately measure the power consumption and the latency of the operations, and the same is true for Design Compiler's modeling of the coreset engine. The simulation tools used in our experiment are widely used and accepted in both industry and research.   We evaluate our simulation using two different datasets, MHEALTH~\cite{mHealth,mHealthDroid}, and
PAMAP2~\cite{PAMAP21,PAMAP22}. The coreset re-construction GAN and DNN models are trained and quantized using tensorflow~\cite{tensorflow}. Furthermore, we also leverage the temporal nature of HAR by designing a dynamic coreset construction algorithm.

\noindent \textbf{\underline{Activity Aware Coreset Construction}:~}
\label{subsubsec:AAC}
In our experiments on HAR sensor data, we observe that not all the activities are equally complex and hence may or may not need a certain number of clusters to represent every feature. While activities like walking and running do not lose much accuracy even when represented with as low as eight clusters, complex activities are more sensitive and need more number of clusters to preserve their geometry. As the communication overhead depends on the number of clusters, which, in turn, depends on the complexity of the activity, we propose an \emph{activity aware clustering} which ensures that coresets for the current activity are represented with just sufficient number of clusters to preserve accuracy. We determine the number of clusters required as a function of current energy availability and accuracy trade off of using a lesser number of clusters. However, naively framed, this approach requires knowledge of what activity is being performed in order to encode the data that will be used to perform inference to determine what activity is being performed. To break this circular dependency, we take inspiration from prior work in HAR~\cite{Origin}, and use the highly stable temporal continuity of human activity (relative to the tens of milliseconds timescales for HAR inferences) to predict the current activity based on previously completed local inferences. We use temporal continuity to our advantage, and make sure that if the system does not have enough energy to form the default 12 clusters, it will resort to forming a smaller number of clusters with minimum accuracy loss. We implement a small lookup table to carry the information on accuracy trade-off for different activities with respect to the number of clusters used to form the coresets (similar to the data represented Fig.~\ref{Fig:AAC-Accuracy}).
We observe that AAC communicates about \textbf{11\%} data compared to sending the full raw data (refer Figure~\ref{Fig:clusters-datavol}). Note that we only resort to activity awareness while forming coresets using clustering, as importance sampling based coreset construction does not require much energy. Furthermore, dropping the number of samples in importance sampling method, even with recovery, significantly hampers the accuracy, which is not the case for recoverable-clustering based coreset construction. 

\noindent \textbf{\underline{Baseline}:~}We choose three points of comparison for our accuracy evaluation. \textbf{Baseline-1 (Baseline: Large DNN)} consists of a full precision (without any pruning) DNN built on the lines of~\cite{HARHaChoi}. \textbf{Baseline-2 (Baseline: EAP)} optimizes (using~\cite{netadapt}) Baseline-1 to design a power-aware DNN tuned to for the average harvested power of our EH source. \textbf{Baseline-3 (Baseline: Origin)} uses the system design proposed in~\cite{Origin}. Baseline-1 and Baseline-2 run on fully powered systems where as Baseline-3 runs on the same EH source as \textit{Seeker}. For communication, we consider a system which transmits the entire raw data to the host as the baseline.   

\noindent \textbf{\underline{Analysis of Results on HAR}:~}Figure~\ref{Fig:MHEALTH-accuracy} and ~\ref{Fig:PAMAP-accuracy} show the accuracy of various policies described in Section~\ref{sec:4}, along with the accuracy of \emph{Seeker} - which applies all policies together, along with ensemble learning. Figure~\ref{Fig:clusters-datavol} shows the normalized data communication volume with different numbers of clusters, along with activity-aware clustering. We make the following observations:

\noindent \textbf{Seeker at Edge Finishes More Work:~}Equipped with multiple decision options, \emph{Seeker} could finish close to 60\% (58.67\% using one of the RE sources; refer Figure~\ref{Fig:Sensitivity-fracCompleted}) of the inferences at the edge itself, thanks to the two efficient and quantized DNNs and the correlation engine. Both the DNNs share the load of the inference depending on the available energy, while correlation engine gets rid of close to 6\% of the redundant compute (refer Figure~\ref{Fig:Sensitivity-fracOffloaded}).  

\noindent \textbf{Seeker at Edge Efficiently Offloads:~}For the unfinished compute, \emph{Seeker} converts the data into coresets for communicating them to the host with minimum payload footprint. The activity aware coresets, thanks to their dynamic nature, reduces the communication volume $8.9\times$ (refer to Figure~\ref{Fig:clusters-datavol}) compared to sending the raw data, and up to $3\times$ compared to the classical compression techniques.

\noindent \textbf{The Recovered Coresets Give Accurate Inference:~}Even with a specialized DNN tranined with coreset data, compressed coresets give less accuracy, evidently because of the loss of features during the coreset formation. However, with the reconstruction (via GAN or cluster redistribution), the accuracy reaches 86.8\% compared to 76.4\% for the former. The GAN modeled the lost signals with a very correlation ($\ge 0.9$ in most cases and $0.6$ in some of the worst cases).

\noindent \textbf{Seeker is Close to a Fully Powered System:~} \emph{Seeker}, thanks to synergistic computation, achieves 87.05\% accuracy with MHEALTH (0.18\% less accurate than a DNN running at full precision with full power) data set and similarly reaches 74.2\% accuracy on the PAMAP2 data set. This gives us $\approx7\%$ more accuracy than~\cite{Origin} on MHEALTH and $\approx3\%$ more accuracy for PAMAP2 dataset. The accuracy improvement is because of three reasons: 1. the DNNs at the edge are more fine-tuned towards delivering accurate results and are $\approx1.5\%$ more accurate than the prior state-of-the-art~\cite{Origin}; 2. the recovered coresets imitate the original data with a great accuracy, and hence the inference accuracy at the host is as good as it would have been with the original data. 3. because \emph{Seeker} could finish more number of inferences (either at the edge or by offloading to a host), greatly reducing the scheduled task from $\approx 40\%$ to $\approx5\%$ in best case, $\approx8\%$ in worst case, and $\approx6.15\%$ in average case (all experiments on RF power trace). To compare with a battery operated energy optimized (to the average energy harvested by the RF sources) system, \emph{Seeker} is $5.89\%$ more accurate on MHEALTH dataset, and $4.98\%$ more accurate on PAMAP2 dataset.

\noindent \textbf{\underline{Sensitivity Study}:~}Our experiments with other EH sources show the versatility of \emph{Seeker}, which outperforms a HAR classifier designed for EH~\cite{Origin} across multiple (piezo-electric, RF) harvested energy sources and demonstrate that it can easily be scaled to work with any number of sensors. Fig.~\ref{Fig:Sensitivity-fracCompleted} shows the comparison of scheduled inferences completed while using \emph{Seeker} and the state-of-the-art~\cite{Origin}. Further, we demonstrate how \emph{Seeker} leverages \textbf{all} the proposed design components (including memoization, DNN inference and coreset) to complete maximum compute at the edge and offload minimum to the host device. Fig.~\ref{Fig:Sensitivity-fracOffloaded} shows the compute breakdown among components under different EH sources.

\begin{figure}
  \centering
  \includegraphics[width=\linewidth]{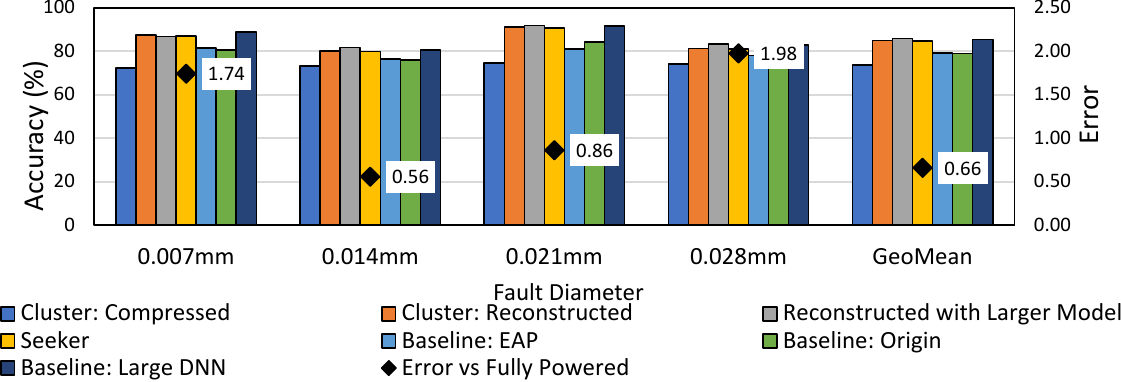}
  \caption{Accuracy of Seeker on Bearing Data set}
  \label{Fig:bearingdata}
 \end{figure}
\vspace{-12pt}
\subsection{\textit{Seeker} for Predictive Maintenance} 
With the advent of automation and industry 4.0~\cite{industry4, industry40}, predictive maintenance is once of the most sorted after problems in industrial IoT domain. Predictive maintenance is a condition-driven preventive maintenance program where, instead of replying on failure to schedule maintenance activities, predictive maintenance uses direct monitoring of the plant to preemptively schedule maintenance and possibly take measures to prevent them from occurring~\cite{predictive}.  For continuous monitoring, the machines are typically fitted with sensors (vibration, force, magnetic, acoustic etc.), and the plant performs continuous analytics on those sensor data for preemptively maintenance scheduling. Vibration based condition monitoring is one of the most common scenarios~\cite{bearing}. Since there are multiple machines, and each machine is fitted with one or more sensors, this is a perfect example of wireless sensor network. The rise of Industry 4.0~\cite{industry4} has lead to an exponential explosion of such sensors in industries, especially in remote or hostile locations, calling for energy harvesting as a solution, and hence need for EH-WSNs. 

\noindent \textbf{\underline{Baseline}:~}Towards this we take Case Western Bearing Fault data set~\cite{bearing} where the vibrations from the different bearings are collected to analyse the fault patterns (e.g. size of a crack in the bearing with respect to operating load, speed etc.). There has been a large body of work~\cite{BearingCNN1,BearingCNN2,BearingCNN3,BearingCNN4} in industrial engineering domain to develop DNN classifer for this task. In our experiments, we took inspirations from the work of ~\cite{BearingCNN4,BearingCNN2} to build a classifier, and applied further optimizations, as we did for HAR evaluation, to make the DNN edge-friendly. We also tweaked AAC to be energy aware only, i.e. the number of clusters formed depends only the energy available. We redesigned (few changes in the hyper parameters) and trained the GAN to adapt to the bearing data for recovering the importance sampling coresets.

\noindent \textbf{\underline{Results}:~}As the base design of \textit{Seeker} can adapt to any sensor based communication, most of the arguments made for HAR still holds true in the case of bearing fault detection. As depicted on Figure~\ref{Fig:clusters-datavol} and~\ref{Fig:Sensitivity-fracCompleted}, \textit{Seeker} reduces the communication overhead by $\approx7\times$ while finishing $\ge 80\%$ of the scheduled compute using WiFi sources. Further, as shown in Figure~\ref{Fig:bearingdata}, Seeker, on an average, delivers an accuracy of $84.73\%$ which is only $0.66\%$ less than a fully powered system. It is note worthy that the bearing data is much susceptible to the real-world nuances and machine part interactions yet, the accuracy of fault prediction is extremely essential towards the production continuity and quality. To portray an example of scale, for a typical grinding job in a manufacturing industry that takes about 8.2 seconds~\cite{grindtime}, and the $5\%$ improvement we observe (in Figure~\ref{Fig:bearingdata}) over the prior state of the art~\cite{Origin} impacts $\approx 46k$ parts per year per machine (working 8 hours/day). Therefore, in large scale industries, both saving communication overhead while maximizing accuracy directly impact the economics of production.  

\vspace{-8pt}

%% file: src/06Conclusion.tex
As systems utilizing energy harvesting edge devices are tasked with increasingly complex tasks, like HAR or predictive maintenance, both system and node designs must respond with targeted efficiency-maximizing optimizations. Our proposal, \emph{Seeker}, synergizes EH sensor nodes host devices by intelligently distributing computations among them, while significantly minimizing the communication overheads. Our experiments show that, by leveraging coreset techniques in data reduction and tuning these techniques for application-aware properties, \emph{Seeker} can reduce the communication overhead by $\approx 8.9\times$, while providing better accuracy ($86.8\%$), even when limited to harvested power, than state-of-the-art energy-optimized DNNs running on a fully powered device ($81.2\%$). Furthermore, it also outperforms the state of the art system designed specifically for EH-WSNs. Collectively, the optimizations in \emph{Seeker}  reduce communication traffic, while improving inference accuracy, demonstrating the potential of holistic system/node/application optimization for the current and future generation of (energy harvesting) wireless sensor nodes.

%% file: src/Appendix.tex
\subsection{Reconstructing Importance Sampling Coreset}
\begin{figure}
  \centering
  \includegraphics[clip,width=\linewidth]{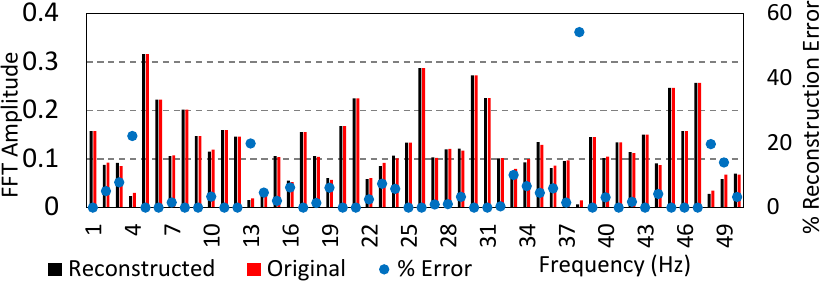}
  \caption{An example of generator based coreset recovery}
  \label{Fig:gen}
 \end{figure}

As discussed in Section~\ref{sec:recover}, we use GANs to recover the data we lost while performing importance sampling. This was motivated from the observation that as we selected more number of points in importance sampling, the accuracy of the inference on the compressed data increased significantly (at times by $2\%$). Hence, the points which were not selected while performing importance sampling still had some importance and can be represented as a function containing the low level nuances of the activity performed and the sensor state. The challenge was to learn this function, i.e. to device a transformation function which can mimic the sensor signal given the aactivity and the sensor states. A similar problem, in terms of generating faces, paintings etc. given some latent space has already been solved using GANs~\cite{olszewski2017realistic}. Motivated by this, we designed a GANs to regenerate the lost data points while performing importance sampling. The latent space takes the activity, and the first and second order moments of the data sample to recreate the signal, and the Discriminator tried to distinguish between the generated signal and the actual signal. The generator is tuned repeatedly until the discriminator could not distinguish the original and the generated signal. The GAN modeled the lost signals with a very high correlation ($\ge 0.9$ in most cases and $0.6$ in some of the worst cases (refer Figure~\ref{Fig:gen} for an example). In rare cases (once in over 2000 cases), the generator induced artifacts which could result in wrong classifications. However, this error could be rectified with further fine tuning. 

\subsection{More Results on Bearing Fault Data}
\begin{figure}
  \centering
  \includegraphics[clip,width=0.8\linewidth]{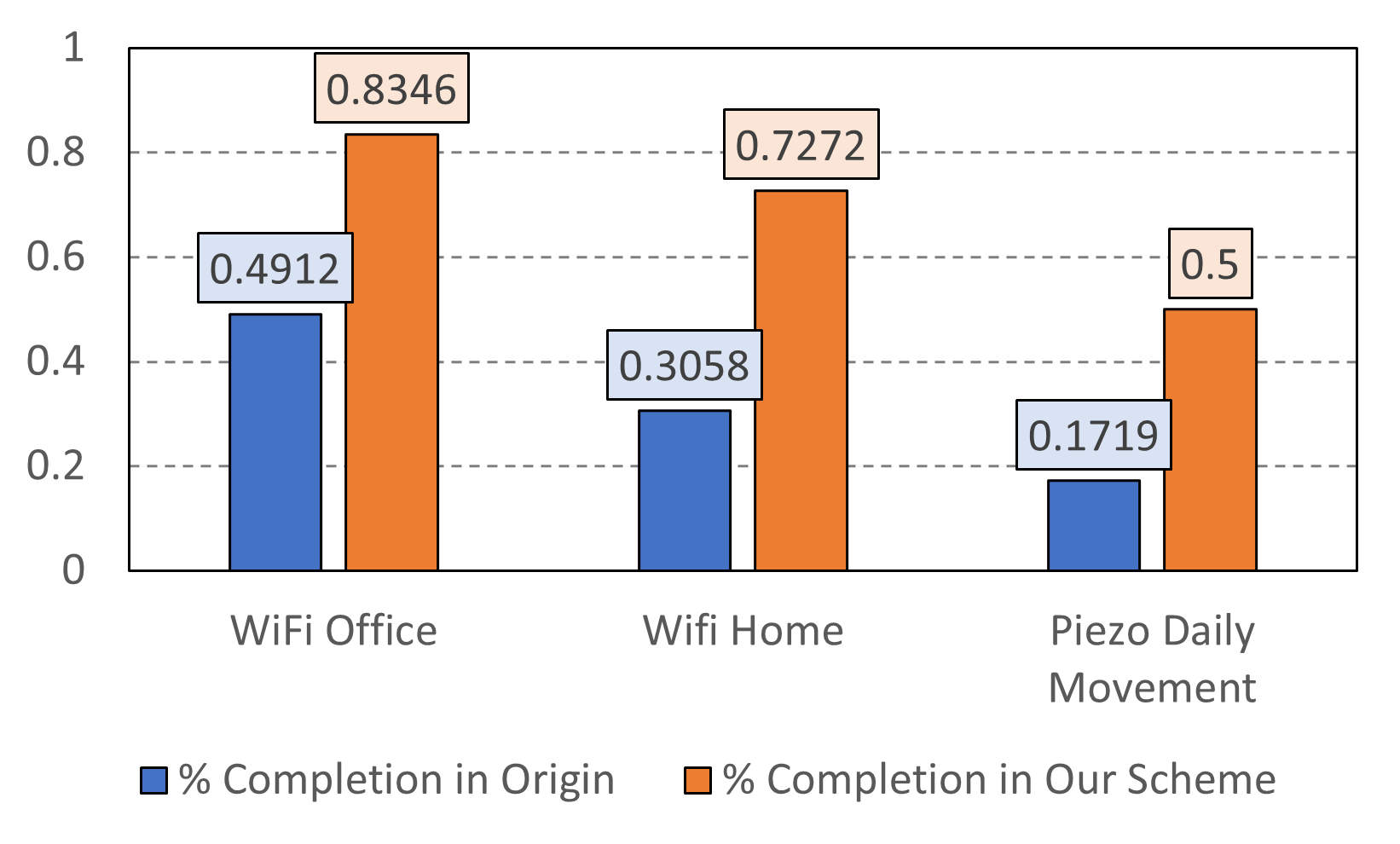}
  \caption{\% completion of the inference at the edge for bearing fault data with different EH source.}
  \label{Fig:bComplete}
 \end{figure}

We repeated our experiments with similar experimental setup on the bearing fault data set~\cite{bearing}. The bearing fault data is sampled at a much higher frequency (48KHz) than the HAR data, and hence require a larger DNN, larger number of importance sampling, and more number of clusters. We took the learning from multiple domain specific literatures~\cite{bearing, eren2019generic, hoang2017convolutional} to isolate the frequency regions specific to the fault pattern to minimize the computations. But, because of the larger data volume, the number of computations performed at the edge diminished significantly (refer Figure~\ref{Fig:bComplete}). We also conducted an empirical study on number of clusters required, and found out that the bearing set data needs about 15 to 20 clusters to maintain the inference accuracy. The data volume communicated for different number of clusters is represented in Figure~\ref{Fig:bearingdata}. 

 \begin{figure}
  \centering
  \includegraphics[clip,width=0.8\linewidth]{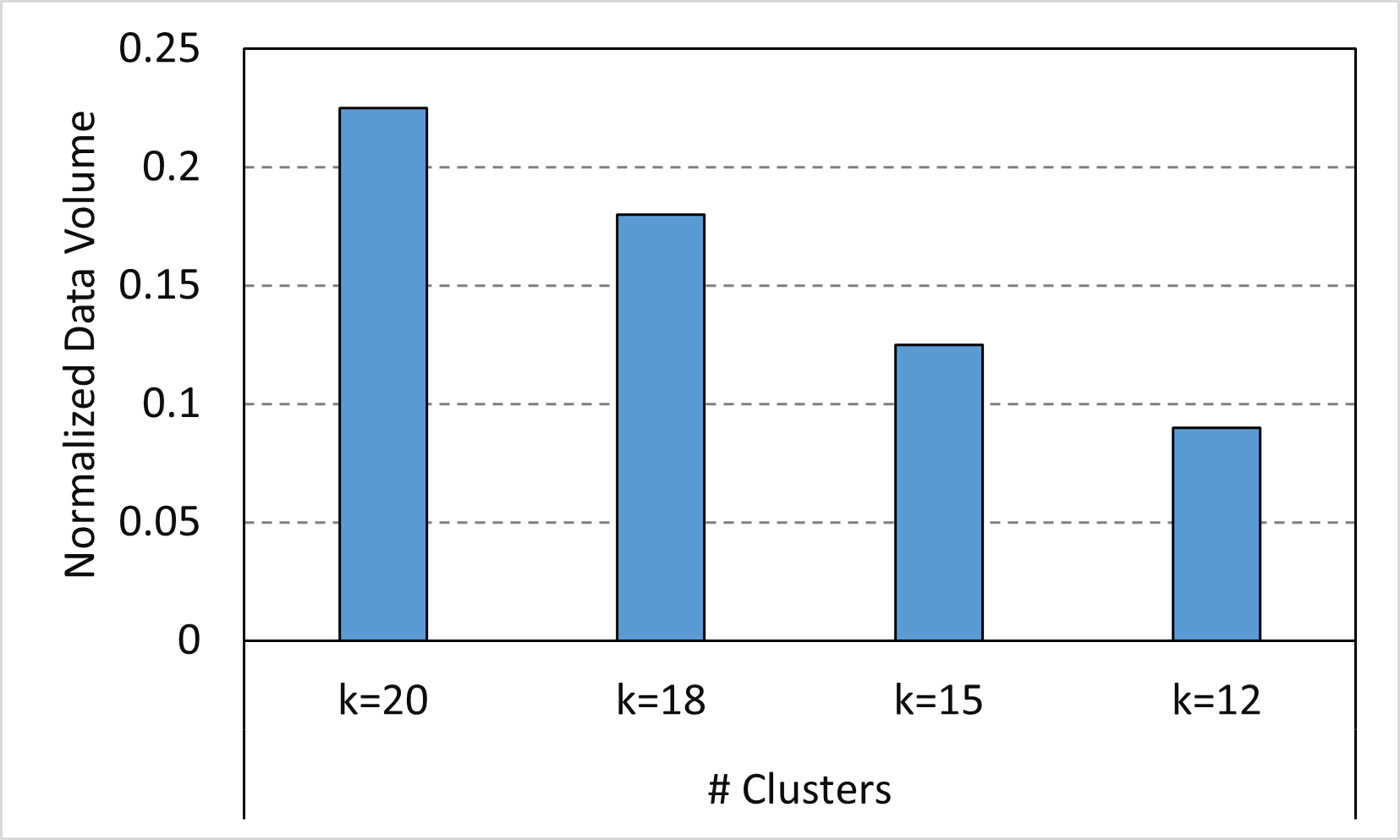}
  \caption{Communication data volume with different number of clusters.}
  \label{Fig:bCluster}
 \end{figure}

 \begin{table}
\centering
\resizebox{\linewidth}{!}{%
\begin{tabular}{|l|c|c|c|} 
\hline
Component & Spec & Power & Area(mm\textsuperscript{2}) \\ 
\hline
SRAM Buffers & \begin{tabular}[c]{@{}c@{}}1kB*256+\\8kB*256+\\64kB+16*256kB\end{tabular} & 10.372W & 117.164 \\ 
\hline
MAC Unit & \begin{tabular}[c]{@{}c@{}}(8*8)\\*256\end{tabular} & 8.46W & 32.72 \\ 
\hline
\begin{tabular}[c]{@{}l@{}}Adder Tree and \\Comparator\end{tabular} & 16*16bit + 256 & 2.4W & 21.556 \\ 
\hline
Control & -- & 0.96W & 12.2 \\ 
\hline
Host & $\sim$Cortex A78 series & 11W & -- \\ 
\hline
\multicolumn{4}{|c|}{Design at 592MHz with~Synopsys AED 32nm library} \\ 
\hline
\multicolumn{1}{|c|}{\textbf{Total}} & 256 tiles & 33.192W & 183.64 \\
\hline
\end{tabular}
}
\vspace{-4pt}\caption{Area and power estimation of our design.}
\label{tab:specs}
\vspace{-14pt}
\end{table}

\begin{figure*}[ht]
\centering
    \subfloat[Accuracy with MHEALTH dataset]
    {
    \includegraphics[width=0.48\linewidth
    , height=2.5cm
    ]{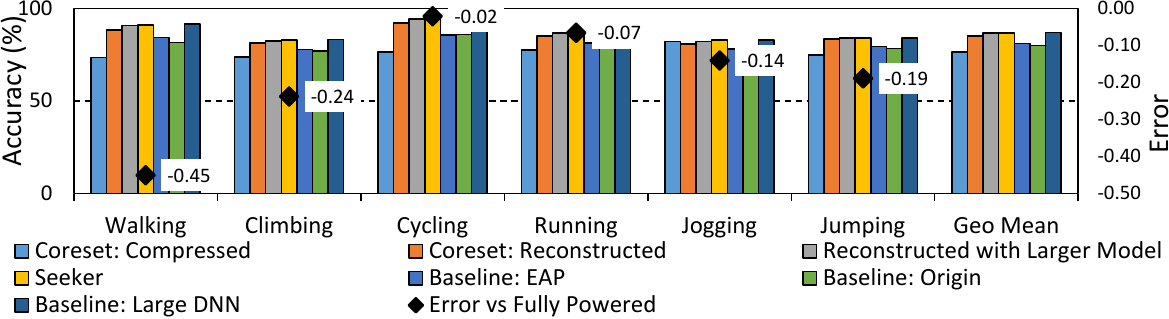}\label{Fig:MHEALTH-accuracy}
    }
    \hfill
    \subfloat[Accuracy with PAMAP2 dataset]
    {
    \includegraphics[width=0.48\linewidth
    , height=2.5cm
    ]{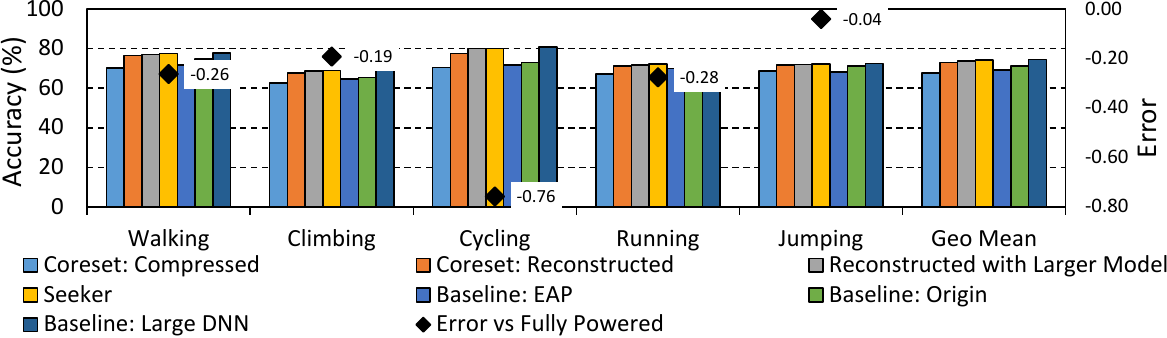}\label{Fig:PAMAP-accuracy}
    }
\caption{Accuracy and communication efficiency of \emph{Seeker} with different data sets and sensitivity study.}
    \label{fig:AllAccuracy}  
\end{figure*}